\documentclass[sn-mathphys-num]{sn-jnl}% Math and Physical Sciences Numbered Reference Style 

\usepackage{graphicx}%
\usepackage{multirow}%
\usepackage{amsthm}%
\usepackage{mathrsfs}%
\usepackage[title]{appendix}%
\usepackage{xcolor}%
\usepackage{textcomp}%
\usepackage{manyfoot}%
\usepackage{booktabs}%
\usepackage{algpseudocode}%
\usepackage{listings}%
\usepackage{textcomp}
\usepackage{pgfplots}
\usepackage{pgfplotstable}
\pgfplotsset{width=10cm,compat=1.9}
\usepackage{filecontents}

\usepackage{framed}
\usepackage{url}
\usepackage[ruled,vlined,linesnumbered]{algorithm2e}
\usepackage{comment}
\usepackage{tikz}
\usetikzlibrary{positioning,quotes}
\usepackage[utf8]{inputenc}
\usepackage{array}
\usepackage{color, colortbl}
\usepackage[T1]{fontenc}
\usepackage{todonotes}
\usepackage{float}
\usepackage{comment}
\usepackage{array, multirow}
\usepackage{soul}
\usepackage{multirow}
\usepackage[export]{adjustbox}
\usepackage{amsfonts} 

\usepackage{lmodern}

\usepackage{lineno}
\usepackage[leqno]{amsmath}
\usepackage{threeparttable}
\usepackage{amsmath}
\usepackage{multirow}
\usetikzlibrary{arrows,shapes}
\usepackage{algorithmicx}
\newcommand{\PBH}{\text{{\scriptsize{LODBH}}}}
\newcommand{\CSH}{\text{{\scriptsize{CSBGH}}}}
\newcommand{\UDH}{\text{{\scriptsize{UDCIM~\cite{working!}}}}}

\newcommand{\getseedsb}{\text{{\scriptsize{GetSeed$S_B$}}}}
\newcommand{\getparent}{\text{{\scriptsize{FindParent}}}}
\newcommand{\tempactive}{\text{{\scriptsize{TemporaryActive}}}}

\newcommand{\PBE}{\text{Local Out-Degree Based Heuristic}}
\newcommand{\CSE}{\text{Community Sized-Base Genetic Algorithm}}

\newcommand\clearrow{\global\let\rowmac\relax}

\newcounter{cases}
\newcounter{subcases}[cases]

\usepackage{geometry}
\usepackage{multirow,tabularx}

\newcolumntype{Y}{>{\centering\arraybackslash}X}

\SetKwComment{Comment}{$\triangleright$\ }{}

\tikzstyle{vertex}=[circle,fill=black!25,minimum size=12pt,inner sep=0pt]
\tikzstyle{selected vertex} = [vertex, fill=red!60]
\tikzstyle{edge} = [draw,thick,-]
 \tikzstyle{weight} = [font=\small]
\tikzstyle{selected edge} = [draw,line width=3pt,-,red!50]
\tikzstyle{ignored edge} = [draw,line width=3pt,-,black!20]

\let\cline\cmidrule

\begin{document}
	
	\title[Community-Based Efficient Algorithms for User-Driven Competitive Influence Maximization in Social Networks]{Community-Based Efficient Algorithms for User-Driven Competitive Influence Maximization in Social Networks}
	
	%%=============================================================%%
	%% GivenName	-> \fnm{Joergen W.}
	%% Particle	-> \spfx{van der} -> surname prefix
	%% FamilyName	-> \sur{Ploeg}
	%% Suffix	-> \sfx{IV}
	%% \author*[1,2]{\fnm{Joergen W.} \spfx{van der} \sur{Ploeg} 
		%%  \sfx{IV}}\email{iauthor@gmail.com}
	%%=============================================================%%
	
	\author*[$^1$]{\fnm{Rahul Kumar} \sur{ Gautam$^{0009-0009-7693-3863}$ }}\email{rvkrgau@gmail.com}
	
	%\author[$^2$]{\fnm{Anjeneya Swami} \sur{Kare$^{0000-0003-3644-4802}$}}\email{askcs@uohyd.ac.in}
	%\equalcont{These authors contributed equally to this work.}
	
	%\author[$^3$]{\fnm{S. Durga} \sur{Bhavani$^{0000-0003-4413-0328}$}}\email{sdbcs@uohyd.ac.in}
	%\equalcont{These authors contributed equally to this work.}
	
	\affil[]{\orgdiv{Department of Computer Engineering and Applications}, \orgname{GLA University}, \orgaddress{\street{Jait}, \city{Mathura}, \postcode{281406}, \state{Uttar Pradesh}, \country{India}}}
	
	%\affil[2]{\orgdiv{Department}, \orgname{Organization}, \orgaddress{\street{Street}, \city{City}, \postcode{10587}, \state{State}, \country{Country}}}
	
	%\affil[3]{\orgdiv{Department}, \orgname{Organization}, \orgaddress{\street{Street}, \city{City}, \postcode{610101}, \state{State}, \country{Country}}}
	
	%%==================================%%
	%% Sample for unstructured abstract %%
	%%==================================%%
	
	\abstract{
     Nowadays, people in the modern world communicate with their friends, relatives, and colleagues through the internet. Persons/nodes and communication/edges among them form a network. Social media networks are a type of network where people share their views with the community. There are several models that capture human behavior, such as a reaction to the information received from friends or relatives. The two fundamental models of information diffusion widely discussed in the social networks are the Independent Cascade Model and the Linear Threshold Model.~\citeauthor{working!}~\cite{working!} propose a variant of the linear threshold model in their paper title User-driven competitive influence Maximization(UDCIM) in social networks. Authors try to simulate human behavior where they do not make a decision immediately after being influenced, but take a pause for a while, and then they make a final decision. They propose the heuristic algorithms and prove the approximation factor under community constraints( The seed vertices belong to an identical community). Even finding the community is itself an NP-hard problem. In this article, we extend the existing work with algorithms and LP-formation of the problem. We also implement and test the LP-formulated equations on small datasets by using the Gurobi Solver~\cite{gurobi}. We furthermore propose one heuristic and one genetic algorithm. The extensive experimentation is carried out on medium to large datasets, and the outcomes of both algorithms are plotted in the results and discussion section.   
    }
\maketitle

\section{Introduction}
With the revolution of the internet, the demand for smartphones has increased significantly.  Via smartphone, a person communicates with friends, colleagues, and relatives through the internet. A person on the internet is considered a node, and communications among them are links/edges that form a social network. Nowadays, people find or post information about basic necessities like buying or selling a product on social media networks. Suppose a person wants to buy a product A and they start looking on internet sources like social media websites.  While searching on a social media platform, the person may be influenced by a new product that may lead them to buy the new product. A company $X$ makes a product $A$,  and company X tries to influence more customers to grow their revenue. The company X finds the highly influential people on social media and asks them to forward our company advertisement so that a larger number of people on the networks can be influenced by their products. The problem is only to find the highly influential people who can maximize the influence of a product/information on social networks.  The propagation of information continues for some periods of time. We call the problem Influence Maximization (IM). 

But, if there is another company $Y$ at the same time, it also wants to influence people to sell their product $B$ , which is similar to $A$. Then the competition arises among the companies that make the same type of products. Suppose we only want to maximize the influence of the product $B$. 

For the above case, if the product information of type $A$ is replaced with fake news and the truth can be replaced with the product information of the type $B$, then the influence of product information of $B$ to be maximized to block the fake information in the social media networks. This problem is known as rumor minimization in  social media networks.

The diffusion models depict the whole process of information diffusion in social media networks. There are two widely accepted models by researchers: The Linear Threshold Model (LTM) and the Independent Cascade Model(ICM).~\citeauthor{kempe2005influential}~\cite{kempe2005influential} propose the Target Set Selection (TSS) problem under the LTM/ICM, which is also known as the influence maximization problem. For the IM problem, we have an input graph $G(V,E)$ ( set of vertices $V$ and set of edges $E$) and the size of the target set $k\in \mathbb{Z^+}$. The objective is to find the target set of size $k$ that maximizes the influence under the  diffusion model, either LTM or ICM. 

Inputs for the LTM are  a graph $G$ and the threshold function $t: V \rightarrow \mathbb{Z^+}$ , where $\mathbb{Z^+}$ is a positive integer. A vertex $v \in V$ is influenced if either it belongs to the target/seed set or $v$ has $t(v)$ active neighbors that are spreaders. The process repeats until the change of state of the nodes from non-active to active continues.

Similarly,  inputs for the ICM are the graph $G$ and the probability function $P: E \rightarrow [0,1]$. The probability function of an edge to $[0,1]$ is the weight on  the edge in $E$. A vertex $u$ is activated or influenced if either $u$ belongs to the target set or any active neighbor $v$ of $u$ can activate $u$ with the probability $P(v,u)$ on the incoming edge $(v,u)\in E$ from the neighbor $v$ to $u$. The process continues until the change in state of the vertex from non-active to active continues. Once a vertex changes its state, it does not revert to its original state.

\citeauthor{working!}~\cite{working!} study the User-driven competitive influence maximization in social networks (UDCIM) and present an approximation algorithm based on community in the given graph. They also propose a variant of the LTM model for the UDCIM problem. The inputs for the UDCIM problem are a graph $G$,  thresholds $\theta_1, \theta_2 \in [0,1]$, seed set $S_A$ of size $k$. The goal is to find the seed set $S_B$ of size $k$ that maximizes the influence of the seed set $S_B$ over $S_A$. A vertex $u$ is activated or influenced if either the vertex $u$ belongs to the seed set $S_A/S_B$ or the sum of weight on incoming active neighbors  of $u$ to $u$ is greater than or equal to either $\theta_1$ or $\theta_1 + \theta_2$. The vertex changes status from non-active to active state by going through two stages: first, temporary activation, and second, final activation state. The process continues till a node changes its state from non-activation to active. The detailed discussion  of the diffusion model of the UDCIM proposed by ~\citeauthor{working!}~\cite{working!} is presented in Section~\ref{sec:prob}. 
\\

Our contributions in the paper are as follows.
\begin{enumerate}
    \item We propose the Linear Programming formulation for the UDCIM problem and also implement it using the Gurobi~\cite{gurobi} framework.
    \item We propose the \PBE{} for the UDCIM problem.
    \item A \CSE{} is also presented based on community size and page rank centrality measure.
    %\item The performance of  our algorithms over recent work is shown by plotting computation time and results graph as shown in \figureautorefname~\ref{fig:inf(A10)} and ~\ref{fig:time(10)}. %We test our algorithms on real-world datasets. 
\end{enumerate}

We discuss the problem in the Section~\ref{sec:prob}, literature review in the Section~\ref{sec:lr} and present the LP-formulation of UDCIM problem in Section~\ref{sec:LP}, propose \PBE{} based on the page rank centrality measure, out-degree, and community based in Section~\ref{sec:a2} and also present a Genetic Algorithm based on community size and page rank centrality measure in the Section~\ref{sec:a3}. In the Section~\ref{sec:result}, we plot the comparison  of the results of the algorithms. Not only the quality of the result, our algorithms take less time as compared to recent work with better results.  The article concludes the paper with the Section~\ref{sec:cnc}.

\section{Recent Work}
\label{sec:lr}

Due to the high use of social media networks by people, researchers show interest in social network analysis. Either we desire to run a social media campaign or do an advertisement through social media. The influence maximization or maximizing the influence of an advertisement is similar to some extent because the objectives of both processes are the same.

~\citeauthor{kempe2003maximizing}~\cite{kempe2003maximizing}  study the problem Target Set Selection (TSS) or Influence Maximization (IM) titled paper Maximizing the spread of influence through a social network. They prove that the IM problem is NP-hard by reducing from Vertex Cover to the IM problem. They also state that the IM problem is hard to approximate within a factor $n^\epsilon$ where $\epsilon>0$.~\citeauthor{chen2009approximability}~\cite{chen2009approximability} also prove in- approximability  of IM problem within linear logarithmic factor.~\citeauthor{kempe2005influential}~\cite{kempe2005influential}  propose the Decreasing Cascade Model. In this model, information flows in the networks  probabilistically.~\citeauthor{kempe2005influential}~\cite{kempe2005influential} also contribute to  the approximation algorithm within the factor $1-1/e-\epsilon))$ on the IM problem under the Decreasing Cascade model, where $e$ is the  natural logarithm base and $\epsilon>0$. The IM problem is Hard to approximate, so seeing empirical results for IM becomes very important. There are several variants of IM problems, and the heuristic results are proposed.~\citeauthor{cordasco2019active}~\cite{cordasco2019active} propose the Perfect Awareness Problem (PAP) under the Linear Threshold Model(LTM). They formulate the problem and propose an LP formulation and heuristic results. Some empirical results on the type of PAP are ~\cite{2023RahulCent,intmax,gautam2023heuristics,2023tieredInf,cordasco2018evangelism}. 

There are a few related problems to the  IM problem, like graph burning~\cite{gautam2022faster,gautam2024approximation}, rumor minimization~\cite{rumor2020containment,rajak2024genetic}, firefighter~\cite{chlebikova2014firefighter}, opinion maximization~\cite{openion2013} and k-center problems~\cite{k-center}. These problems take different inputs, but the objective of finding potential nodes is to cover/maximize influence on more vertices. For example, if we talk about rumor minimization~\cite{rumor2020containment}, we set goal to stop diminishing the rumor in the social networks. To counter the rumor, we have to spread the truth to convince more people so they stop relying on rumors. Here, the truth maximization is our goal, like influence maximization. Similarly, in the Graph burning problem, the selection of potential nodes is crucial so that all together they can burn the maximum number of nodes. The differences are only in the diffusion processes of information with the same or a slightly different objective.

In the IM or TSS problem, at a time, only one information propagates and influence is maximized.  But in a competitive world, there may be more than one information propagating simultaneously, and sources of information want to increase their influence individually.~\citeauthor{bharathi2007competitive}~\cite{bharathi2007competitive} propose the approximation algorithm with factor $1-1/e$ where $e$ is natural logarithm base for the competitive influence maximization problem. They also propose an FPTAS for the bi-directional tree graphs for maximizing the influence of a single player.~\citeauthor{2023targetedCompt}~\cite{2023targetedCompt} formulate  the Target Influence Maximization in Competitive Social networks (TIMC) and propose the heuristic algorithms under the Independent Cascade Model(ICM).~\citeauthor{working!}~\cite{working!} propose the UDCIM problem with the consideration of human behavior, as human takes time to make a decision after being influenced by the information. In this paper, we propose the LP-formulation of UDCIM, two heuristics, and the implementation of the LP-formulation on small datasets by using Gurobi~\cite{gurobi}, as we know the UDCIM problem is NP-Hard.

\section{Problem Definition}
\label{sec:prob}
We have a graph $G(V,E)$ with a set of vertices $V$ and a set of edges $E$. Weight on each edge is defined as $W_{(u,v)} \in [0,1]$, where $(u,v)\in E$. There are two competitors $A$ and $B$ want to maximize their influence in the network $G$. As we see in real life, people may have a tendency toward a particular product ($A$ and $B$ types) or may be neutral for both types of information. Initially, each vertex of the graph is  declared either $A$, $B$, or neutral type. A vertex $u_\alpha$ likes and believes in the product $A$ types of information. Similarly, a vertex $u_\beta$  likes and believes in the product $B$ types of information. Remaining neutral vertices are $\gamma$ types and denoted as $u_\gamma$. The objective is to maximize the influenced vertices by one of the competitors after following the diffusion process given below. In this paper, particularly, we maximize the influence of $S_B$  for given $S_A$.

\subsection{Diffusion Process}
We have the seed set $S_A$. The vertices $u\in S_A$ are the initial spreaders of the information type $A$. A vertex $u$ becomes activated if $u$ is temporary activated. So, each vertex goes through two states to be influenced, temporary activated, and then final activation. We explain one by one these two stats of a vertex. 
\subsubsection{Temporary Activation  State}
A vertex $u_\alpha \in V$ is temporary activated if the sum of weight on incoming edges from active neighbors of type $A$ through $u_\alpha$ receives information of type $A$ is greater than or equal to $\theta_1\in [0,1]$, or if sum of weight on incoming edges from active neighbors of type $B$ through $u_\alpha$ receives information of type $B$ is greater than or equal to $\theta_1 + \theta_2$ ($\theta_1,\theta_2 \in [0,1]$). $N[u_\alpha]$ is set of incoming vertices to $u_\alpha$.

$$ \sum\limits_{v_\alpha \in N[u_\alpha]} W_{(v_\alpha,u_\alpha)} \ge \theta_1$$
$$OR$$
$$ \sum\limits_{v_\beta \in N[u_\alpha]} W_{(v_\beta,u_\alpha)} \ge \theta_1 + \theta_2 $$

Similarly, a vertex $u_\beta \in V$ is temporary activated if the sum of the weight on the incoming edge through $u_\beta$ receives information of type $B$ is greater than or equal to $\theta_1\in [0,1]$ , or if the sum of the weight through $u_\beta$ receives information of type $A$ is greater than or equal to $\theta_1 + \theta_2$.

$$ \sum\limits_{v_\beta \in N[u_\beta]} W_{(v_\beta,u_\beta)} \ge \theta_1$$
$$OR$$
$$ \sum\limits_{v_\alpha \in N[u_\beta]} W_{(v_\alpha,u_\beta)} \ge \theta_1 + \theta_2 $$

For a neutral vertex $u_\gamma$ , and is temporary activated if $u_\gamma$ holds condition given below.
$$ \sum\limits_{v \in N[u_\gamma]} W_{(v,u_\gamma)} \ge \theta_1, \; \text{ Where $v$ is either $A$ or $B$ type.}$$

\subsubsection{Final Activation  State}
As each vertex $u \in V$ is at temporary activation state for either $A$ or $B$'s products. It simulates a real-life scenario where a person waits and takes time to decide whether they buy a product. A person buys a product under the final activation. So, a temporary activated vertex $u_\alpha$ is finally influenced for product type $B$ if the vertex $u_\alpha$ satisfies the following condition, as given below. Otherwise, $u_\alpha$ it influences for $A$ and buy $A$ product. 
$$ \sum\limits_{v_\beta \in N[u_\alpha]} W_{(v_\beta,u_\alpha)} \ge max(\sum\limits_{v_\alpha \in N[u_\alpha]} W_{(v_\alpha,u_\alpha)},\theta_1+\theta_2) \; \label{eq:max1cond}$$

Similarly, a temporary activated vertex $u_\beta$ is finally influenced for $B$ if the vertex $u_\beta$ satisfies the following condition as given below. Otherwise, $u_\beta$ influences for $B$ and buy  the product $B$. 
$$ \sum\limits_{v_\alpha \in N[u_\beta]} W_{(v_\alpha,u_\beta)} \ge max(\sum\limits_{v_\beta \in N[u_\beta]} W_{(v_\beta,u_\beta)},\theta_1+\theta_2) \; \label{eq:max1cond2}$$

Remaining for the vertices $u_\gamma \in V$ is finally influenced for $A$, if 

$$ \sum\limits_{v_\alpha \in N[u_\alpha]} W_{(v_\alpha,u_\alpha)} \ge \sum\limits_{v_\beta \in N[u_\beta]} W_{(v_\beta,u_\beta)} \; \label{eq:max1cond3}$$ Otherwise, $u_\gamma \in V$ is influenced for $B$ and buy the product of type $B$.
In the next section, we present the Linear Formulation of the UDCIM problem.
    
\section{LP-formulation}
\label{sec:LP}
The objective of the UDCIM problem is to maximize the number of activated vertices for the product of type $B$ as given in equation  ~\ref{eq:1}. The diffusion process is an iterative process that ends up with $n=|V|$ rounds. For all equations, $r$ belongs to $[0,|V|]$. $B_{u,r}=1$ If the vertex $u$ is active for the type $B$  till $r^{th}$ rounds, otherwise $B_{u,r}=0$. Similarly for  $r\le n$, $A_{u,r}=1$ if the vertex $u$ is active of type $A$ till $r^{th}$ rounds, otherwise $A_{u,r}=0$. To avoid complexity in equations, we define new notation $K(\xi, \alpha, u)$, which is the sum of weight on the incoming edge through $v \in N[u]$,  where $u$ is of $a$ type vertex and $u$ receives information of type $\xi$ ($\xi \in \{A, B\}$) till the $r-1^{th}$ rounds. As shown in the equation~\ref{eq:dfn}, $K(\xi, a, u)$  can be written as.

$$K(\xi,a, u) = \sum\limits_{v\in N[u]} W_{v,u}*\xi_{v,r-1} \label{eq:dfn}$$

\subsection{Temporary Activation: Equations~\ref{eq:t1}-\ref{eq:tb.3}}
 $S_A$ is a seed set to influence the vertices for $A$ types. $S_A$ is already a known set. To activate all the vertices $u \in S_A$ at the beginning, we set  $A_{u,r}=1$ and $B_{u,r}=0$ in ~\ref{eq:t1}. Equations ~\ref{eq:t2} and ~\ref{eq:t3} put restrictions on initial spreaders of each type that can not be more than the given size as $|S_A|$.\\
\textbf{ If $u$ is of $A$ type or $\alpha$ type : Equations ~\ref{eq:ta1}-~\ref{eq:ta.3}}\\
$T^1_{u,r}=1$ if  $K(A,\alpha, u) \ge \theta_1$ and $T^2_{u,r}=1$ if  $K(B,\alpha, u) \ge \theta_1 + \theta_2$ otherwise; $T^1_{u,r}=0$ and  $T^2_{u,r}=0$. $T_{u,r}=1$ if at least one of $T^1_{u,r}$ and $T^2_{u,r}$ is one otherwise $T_{u,r}=0$ . The questions ~\ref{eq:ta.1} to ~\ref{eq:ta.3} ensure the condition. The equation ~\ref{eq:ta.1} forces $T_{u,r}=0$ when $T^1_{u,r}=0$ and $T^2_{u,r}=0$. The equations ~\ref{eq:ta.2} and  ~\ref{eq:ta.3} make $T^1_{u,r}=1$ if one or both $T^1_{u,r}$ and $T^2_{u,r}$ are one.  \\
\textbf{ If $u$ is of $B$ type or $\beta$ type : Equations ~\ref{eq:tb1}-~\ref{eq:tb.3}}\\
Similarly, the equations ~\ref{eq:tb1} to ~\ref{eq:tb.3} all together do the same things as explained for $A$ type.\\ 
\textbf{ If $u$ is of neutral type or $\gamma$ type : Equations ~\ref{eq:tg1}}\\
$T_{u,r}=1$ if $K(B,\gamma,u) + K(A,\gamma,u) \ge \theta_1$ is true; otherwise $T_{u,r}=0$ .

\subsection{Final Activation: Equations~\ref{eq:da1}-\ref{eq:dg.1}}

As we know, temporarily activated vertices are finally activated in the diffusion process. We divide vertices into three categories for activation $A$ types, $B$ types, and $\gamma$ types.\\
\textbf{ If $u$ is of $A$ type or $\alpha$ type: Equations ~\ref{eq:da1}-~\ref{eq:da2}}\\
$B_{u,r}=1$ if  $K(B,\alpha, u) \ge K(A,\alpha, u) $ and $K(B,\alpha, u) \ge \theta_1 + \theta_2 $ or in other words $K(B,\alpha, u)$ is greater than or equal to max of $K(A,\alpha, u)$ and   $\theta_1 + \theta_2 $. otherwise; $B_{u,r}=1$.\\
\textbf{ If $u$ is of $B$ type or $\beta$ type: Equations ~\ref{eq:db1}-~\ref{eq:db2}}\\
Similarly, the equations ~\ref{eq:db1} to ~\ref{eq:db2} all together do the same things.\\ 
\textbf{ If $u$ is neutral type or $\gamma$ type : Equation ~\ref{eq:dg1}}\\
We assign $B_{u,r}=1$ if $K(B,\gamma,u) \ge K(A,\gamma,u)$ is true; otherwise $B_{u,r}=0$.

The equation~\ref{eq:dg.1} serves the condition where $T_{u,r}$ must be temporary activated for final activation as $A_{u,r}=1$ or $B_{u,r}=1$. If $T_{u,r}=0$ then $A_{u,r}=0$ and $B_{u,r}=0$. 

%the vertex $v$ is $\alpha$ types and the condition is  satisfied then the vertex is activated for $B$ types else $A$ types. The equation \ref{eq:2.8} forces $A^3_v=0$ if $B^1_v =1$ and vice-versa. Similarly, The equations ~\ref{eq:2.6} and ~\ref{eq:2.7}  are combined to check condition( $max(\sum\limits_{v_\beta \in N[u_\beta]} W_{(v_\beta,u_\beta)},\theta_1+\theta_2)$) given in the diffusion section ~\ref{eq:max1cond1}.

%If the vertex is of $\beta$ types and the condition is  satisfied then the vertex is influenced for $A$ types else $B$ types. The equations \ref{eq:2.9} force $B^3_v=0$ if $A^1_v =1$ and vice-versa. $A^2_v =1$ if it satisfied the condition given in equation \ref{eq:max1cond3} otherwise $A^2_v =0$ and $B^2_v =1$. In the last, finally check if any $ a_v \in \{ A^1_v, A^2_v,  A^3_v\}$  is one then $A_{(v,r)}=1$. Similarly, if any $b_v \in \{ B^1_v, B^2_v,  B^3_v\}$ is one then $B_{(v,r)}=1$.   

\begin{comment}
    
$$K(A,\alpha)= (\sum\limits_{v \in N[u]} W_{(v,u)}*A_{(v,r-1)} )* \alpha_u$$
$$K(A,\beta)= (\sum\limits_{v \in N[u]} W_{(v,u)}*B_{(v,r-1)} )* \beta_u$$
Similarly,
$$K(B,\beta)= (\sum\limits_{v \in N[u]} W_{(v,u)}*B_{(v,r-1)} )* \beta_u$$
$$K(B,\alpha)= (\sum\limits_{v \in N[u]} W_{(v,u)}*B_{(v,r-1)} )* \alpha_u$$

\end{comment}

\begin{align}
		Objective: \text{Maximize}  \displaystyle\sum\limits_{ u\in V \backslash S_A } B_{u,n} \text{ and } \displaystyle\sum\limits_{ u\in V } A_{u,n} & & \text{ Priority of maximization is for $B_{u,n}$.} \label{eq:1}\\
		\text{subject to the constraints} \\
        \text{Temporary Activation;} \\
        %\textbf{Temporary Activated}
        A_{u,r}=1 \text{ and } B_{r,0}=A_{u,r}-1,  & & \forall u \in S_A, \text{ $r\in [0,|V|]$} \label{eq:t1}\\
        \sum\limits_{u\in  V } A_{u,0} \le |S_A|,  & &  \label{eq:t2}\\
        \sum\limits_{u\in V \backslash S_A} B_{u,0} \le |S_A|,  & &  \label{eq:t3}\\
        K(A,\alpha,u)\ge \theta_1* T^1_{u,r},  & & \text{ If u is of $A$ type}  \label{eq:ta1}\\
         K(B,\alpha,u)\ge (\theta_1+ \theta_2)* T^2_{u,r},  & &  \text{ If u is of $A$ type} \label{eq:ta2}\\
         T_{u,r}\le T^1_{u,r}+T^2_{u,r} & & \label{eq:ta.1}\\
         T_{u,r}\ge T^1_{u,r}  & & \label{eq:ta.2}\\
         T_{u,r}\ge T^2_{u,r} & & \label{eq:ta.3}\\
         K(B,\beta,u) \ge \theta_1 * T^3_{u,r},,  & & \text{ If u is  of $B$ type}  \label{eq:tb1}\\
         K(A,\beta,u) \ge (\theta_1 \theta_2) * T^4_{u,r},,  & & \text{ If u is  of $B$ type}  \label{eq:tb2}\\
         T_{u,r}\le T^3_{u,r}+T^4_{u,r} & & \label{eq:tb.1}\\
         T_{u,r}\ge T^3_{u,r}  & & \label{eq:tb.2}\\
         T_{u,r}\ge T^4_{u,r} & &\label{eq:tb.3}\\
         K(B,\gamma,u) + K(A,\gamma,u) \ge T_{u,r}*\theta_1 & & \text{ If $u$ is neutral/$\gamma$ type}  \label{eq:tg1}\\ 
        \text{Diffusion:}\\
        K(B,\alpha,u)*T_{u,r}\ge K(A,\alpha,u) * (B_{u,r}-B_{u,0})  & & \text{ If $u$ is of $A$ type} \label{eq:da1}\\
        K(B,\alpha,u)*T_{u,r}\ge (\theta_1+\theta_2) * (B_{u,r}-B_{u,0}) ,  & & \text{ If u is of $A $ type}  \label{eq:da2}\\
        K(A,\beta,u)*T_{u,r}\ge K(B,\beta,u)* (A_{u,r}-A_{u,0})   & & \text{ If u is of $B $ type}  \label{eq:db1}\\
        K(A,\beta,u)*T_{u,r}\ge (\theta_1+\theta_2)* (A_{u,r}-A_{u,0}),  & &  \text{ If u is of $B$ type} \label{eq:db2}\\
        K(A,\gamma,u)*T_{u,r}\ge K(B,\gamma,u)* (B_{u,r}-B_{u,0}),  & & \text{ If u is of neutral/$\gamma$ type}  \label{eq:dg1}\\
        A_{u,r}+ B_{u,r} = T_{u,r} & & \label{eq:dg.1}
\end{align}

 The variables hold binary values as $A_{u,r}, B_{u,r}, T_{u,r}\in\{0,1\}$. 
 \subsection{Implementation of LP-Formulation}
 We implement the LP formulation of the UDCIM problem by using Gurobi~\cite{gurobi}. We run the implemented LP-formulation for a small-sized dataset up to $10$ minutes and capture the results. The results are plotted in the \tableautorefname~\ref{tab:gurobi}.

 \begin{table}[!htb]
     \centering
     \begin{tabular}{ccccc}
     \hline
          \multirow{2}{*}{Algorithm Name} &  \multicolumn{2}{c}{$|S_A|=30$}  & \multicolumn{2}{c}{$|S_A|=40$}\\
          \cline{2-5}
          &$\sigma(S_A)$  & $\sigma(S_B)$ & $\sigma(S_A)$ & $\sigma(S_B)$ \\
          \hline
          UDCIM~\cite{working!}&125  & 239 & 133 & 202 \\
         \PBH{} &  15 & 370 & 8 & 356 \\
          Gurobi-Solver&  31&  \textbf{430} &  40 &  \textbf{427} \\
          \hline
     \end{tabular}
     \caption{Implementation of LP-formulation by using Gurobi~\cite{gurobi} on dataset \textit{web-polblogs}~\cite{neworkRepository} of size $643$ vertices and $2280$ edges.}
     \label{tab:gurobi}
 \end{table}

\section{\PBE{}}
\label{sec:a2}
We have a directed graph $G(V,E)$. The algorithm ~\ref{algo:Potential} \getseedsb{} finds the highly potential set of vertices  $S_B$  that can influence a high number of vertices. The algorithm searches the seed vertices in the community of the graph $G$. There may be more than one community in the graph $G$. In a few communities, \getseedsb{} finds the highly potential vertices that can maximize the influence. The highly potential vertices are stored in a list $\hat{P}$, which is in decreasing order of their out degree. The first $k$ set of vertices of the list  $\hat{P}$ is selected as a set of seed vertices $S_B$. Later, the diffusion function returns the set of influenced/activated vertices  $\sigma(S_A)$ and $\sigma(S_B)$ by $S_A$ and $S_B$, respectively. 

The algorithm~\ref{algo:Potential} finds the set $S_B$ of sizes $k$ for a given set $S_A$ of size  $k$. Let $\sigma(S_A)$ be the set of influenced vertices by the seed set $S_A$ of given size $k$, and let $C= \{C_1,C_2,C_3,\cdots, C_m\}$ be the set of communities in the graph $G$ calculated by \textit{Luvian Algorithm}. For each $u\in \sigma(S_A)$, if $u$ belongs to a community $c_i$, then $u$ is added to a temporary set $T_i$. As shown in the algorithm~\ref{algo:findParent}, the \getparent{} function returns an array of all the vertices $v \in C_i$ if there exists a path from $v$ to $u\in T_i$. Let the returned array be called the potential list $\hat{P}$. Sort the list $\hat{P}$ in decreasing order of out-degree of the vertices.  The first $k$  set of vertices in the list  $\hat{P}$ is selected as seed vertices as the seed set $S_B$.     

As shown in Algorithm~\ref{algo:findParent}, the \getparent{} uses the queue data structure to find the vertices reachable from the vertices in the community $C_i$ to $T_i$. Let $Q$ be the empty queue. Initialize a variable the list  $\hat{P}$ with $\phi$. The vertex $u\in T_i$ is enqueued to $Q$ and $u$ is marked visited once the vertex gets out of the queue. We repeat following steps until $Q$ is not empty: If the queue $Q$ is not empty, then dequeue the vertex $u$ from the queue $Q$ and add all vertices $v$ in the list $\hat{P}$ if $v$ is not visited and $v \in N_{in}[u]$ where  $N_{in}[u]$ is the set of vertices associated with incoming edges toward $u$.

\subsection{Diffusion Algorithm}

As shown in the Algorithm~\ref{algo:diffu}, The Diffusion algorithm takes inputs $\theta_1, \theta_2, G, S_A$ and $S_B$. As per the given diffusion model in section~\ref{sec:prob}, to become a final active state, a vertex $u$ has to first be temporarily active. The diffusion function finds all vertices that are temporary active by $S_A$ and $S_B$ sets and to be finally activated. As we have implemented the algorithm in Python, we define a few notations as follows. 
%A $Flag$ variable will take care any change of status of vertex in any round or loop. Initially, $Flag$ is set to $True$.

\begin{align*}   
 G.nodes[u]['tend']=\biggl\{^{1 \text{ ; If u is of $A$ type or $\alpha$ types.}}_{2 \text{ ; If u is of $B$ type or $\beta$ types.}}\\
 G.nodes[u]['tend'] = 0 \text{ if $u$ is of neutral type or $\gamma$ types.}
\end{align*}

\begin{align*}   
 G.nodes[u]['tend']=\biggl\{^{1.5 \text{ ; If $u$ is at final active state of $A$ types.}}_{2.5 \text{ ; If $u$ is at final active state of $B$ types.}}
\end{align*}

$G.nodes[u]['tend']$ tells us what type of $u$ is initially defined or stored in the input dataset.

As we know, an active vertex has to be first in the temporary active state to get into the final activation state. The diffusion function as shown in Algorithm~\ref{algo:diffu}, first calls the \tempactive{} function as shown in Algorithm~\ref{algo:tempAct} then check conditions to be a final activation and if a vertex $v$ is activated for $A$ types then assign value  $1.5$ to $G.nodes[u]['tend']$ and for $B$ types assign value $2.5$ to $G.nodes[u]['tend']$. The while loop runs until a change in the trend of the vertex takes place. At the end, the diffusion function returns the graph $G$ with the updated trend.

\subsection{Temporary Activation}

The Algorithm~\ref{algo:tempAct} (\tempactive{}) function finds the vertices that may be active during the diffusion process. The algorithm takes the inputs $\theta_1, \theta_2, G, S_A, \text{ and } S_B$.  A variable $t_a$ is the sum of  weight on incoming edges to $u$ from $v \in N_{in}[u]$ ($v$ is at the final active state vertex of $A$ types or $G.nodes[v]['tend']=1.5$). Similarly, another variable $t_b$ is the sum of weight on incoming edges to $u$ from $v \in N_{in}[u]$ ($v$ is at the final active state vertex for $B$ types or $G.nodes[v]['tend']=2.5$). The variable $t_c$ is the sum of $t_a$ and $t_b$. Initialize $t_a, t_b$ and $t_c$ by zero. The vertex $u$ is temporary activated and added to the set $T_A$ if the vertex $u$ satisfies the following conditions;

\begin{enumerate}
    \item If $G.nodes[u]['tend']=2$ and $t_b \ge \theta_1$ or $t_a \ge \theta_1 + \theta_2$. 
    \item If $G.nodes[u]['tend']=1$ and $t_a \ge \theta_1$ or $t_b \ge \theta_1 + \theta_2$.
    \item If $G.nodes[u]['tend']=0$ and $t_c \ge \theta_1$.
\end{enumerate}

The \tempactive{} function returns the set of activated vertices $T_A$.

\subsection{Final Activation}
%Let a set $T_A$ is the set of temporary activated vertices.
We have a set of temporary activated vertices $T_A$ computed by Algorithm~\ref{algo:tempAct}.
For each vertex $u \in T_A$, $G.nodes[u]['tend']$ is temporary activated either of $A$ or $B$ types. $t_a$ is the sum of weight on the incoming edge from $v \in N_{in}[u]$ (where $v$ is at final active state of $A$ type or $G.nodes[v]['tend']=1.5$) to $u$. Similarly, $t_b$ is the sum of weight on the incoming edge from $v \in N_{in}[u]$ (where $v$ is at the final active state of $B$ type or $G.nodes[v]['tend']=2.5$) to $u$. There are three situations for final activation due to the types of vertex $u$.
\begin{enumerate}
    \item If $G.nodes[u]['tend']= 1$ and $t_b \ge max(t_a, \theta_1+\theta_2)$, then $u$ is active for $B$ and update $G.nodes[u]['tend']$ to $2.5$  otherwise; active for $A$ and update $G.nodes[u]['tend']$ to $1.5$ type. 
    \item If $G.nodes[u]['tend']= 2$ and $ta \ge max(t_a, \theta_1+\theta_2)$, then $u$ is active for $A$ and update $G.nodes[u]['tend']$ to $1.5$ otherwise; active for $B$ type and update $G.nodes[u]['tend']$ to $2.5$.
    \item If $G.nodes[u]['tend']= 0$ and $tb > ta $, then $u$ is active for $B$ and update $G.nodes[u]['tend']$ to $2.5$ otherwise; active for $A$ type and update $G.nodes[u]['tend']$ to $1.5$.
\end{enumerate}

If at least one vertex changes status, we repeat the diffusion process otherwise we stop it and return the graph $G$.

\begin{algorithm}
	\DontPrintSemicolon
	\SetKwInOut{Input}{Input}\SetKwInOut{Output}{Output}
	\Input   {Graph :$G=(V,E),\text{ thresholds : }\theta_1, \text{and }\theta_2, \text{ seed sets : } S_A, \text{ and } S_B$.}
	\Output  {The seed set $S_B$.}
	\SetKwFunction{FMain}{\getseedsb{}}
	\SetKwProg{Fn}{Function}{}
	\Fn{\FMain{$G, \theta_1,\theta_2, S_A,S_B$ }} {
		\Begin{
                $\hat{P}\gets \phi$\;
                $\sigma(S_A), \sigma(S_B)\gets Diffusion(\theta_1,\theta_2,S_A,S_B,G)$\;
                $C \gets LuvianCommunites(G)$\;
                \For{$u \in \sigma(S_A)$}{
                \For{$ i=1$ to $|C|$}{
                  \If{$u \in C_i$}
                  {
                        $T_i \gets T_i \cup \{u\}$\;
                        break\;
                  
                  }
                }
                }
                \For{$ i=1$ to $|C|$}{
                    \For{$v \in \getparent{}(T_i,G,C_i,S_A)$}{
                        $\hat{P} \gets \hat{P} \cup \{v\}$\;
                }
                }
                $\hat{P} \gets sort(\hat{P})$\Comment{Decreasing order of out-degree of the vertices in list $\hat{P}$ and update itself.}
                $S_B \gets \hat{P}[0:k-1]$ \Comment{Take first $k$ elements of $\hat{P}$ as set of seeds for influence $B$ types of information.}
                %$inf_A, inf_B\gets Diffusion(\theta_1,\theta_2,S_A,S_B,G)$\;
                \KwRet{$S_B$}\;
                
		}
	}
	\caption{Computing the seed set $S_B$}
	\label{algo:Potential}
\end{algorithm}

\begin{algorithm}
	\DontPrintSemicolon
	\SetKwInOut{Input}{Input}\SetKwInOut{Output}{Output}
	\Input   {$G=(V,E) $, $T_i,\text{community set }C_i, \text{and seed set } S_A)$.}
	\Output  {Potential seed set $\hat{P}$.}
	\SetKwFunction{FMain}{\getparent{}}
	\SetKwProg{Fn}{Function}{}
	\Fn{\FMain{$T_i,G,C_i,S_A)$}} {
		\Begin{
                $\hat{P} \gets \phi$\;
                $Q \gets \phi$\;
                \For{$u \in T_i$}{
                    $Q.enqueue(u)$\;
                }

                \While{$Q.empty()=False$}{
                     $u \gets Q.dequeue()$\;
                     $\hat{P} \gets vis \cup \{u\}$\;
                     \For{$v \in N_{in}[u]$}{
                        \If{ $v \in C_i$ and $v \notin \hat{P} \;\cup\ S_A$}{
                        $Q.enqueue(v)$\;
                        $\hat{P} \gets \hat{P} \cup \{v\}$\;
                        }
                     
                     }
                     }
                     
                \KwRet{$\hat{P}$}
                
		}
	}
	\caption{Find all parents of the vertices in $T_i$.}
	\label{algo:findParent}
\end{algorithm}

\begin{algorithm}
	\DontPrintSemicolon
	\SetKwInOut{Input}{Input}\SetKwInOut{Output}{Output}
	\Input   {$\theta_1, \theta_2, G, S_A, \text{ and }S_B$.}
	\Output  {Updated graph with activated vertices.}
	\SetKwFunction{FMain}{Diffusion}
	\SetKwProg{Fn}{Function}{}
	\Fn{\FMain{$\theta_1, \theta_2, G, S_A, S_B$}} {
		\Begin{
                $flag \gets True$\;
                $A \gets \phi$\;
                \While{$falg =True$}{
                $flag \gets False$\;
                  $T_A \gets \tempactive{}(\theta_1, \theta_2, G, S_A, S_B)$\;
                  \For{$u \in T_A $}{
                  $t_a,t_b \gets 0$\;
                  %$status \gets G.nodes[u]['tend]$\;
                  \For{$v \in N_{in}[u]$}{
                  \If{$G.nodes[v]['tend]= 1.5$}{
                      $t_a \gets t_a + W_{(u,v)}$\;
                  }
                  \ElseIf{$G.nodes[v]['tend]= 2.5$}{
                  $t_b \gets t_b + W_{(u,v)}$\;
                  }
                  }
                    
                  \If{$G.nodes[u]['tend']=1$}{
                  $Flag \gets True$\;
                  $G.nodes[u]['tend'] \gets 1.5$\;
                  \If{$tb>= max(t_a,\theta_1+\theta_2)$}{
                   $G.nodes[u]['tend'] \gets 2.5$\;
                  }
                  }

                  \If {$G.nodes[u]['tend']=2$}{
                $Flag \gets True$\;
                $G.nodes[u]['tend'] \gets 2.5$\;
                \If{$ta>= max(tb,\theta_1+\theta_2)$}{
                    $G.nodes[u]['tend'] \gets 1.5$\;
                }
                }
                        
            \If{$G.nodes[u]['tend']=0$}{
                $Flag \gets True$\;
                $G.nodes[u]['tend'] \gets 1.5$\;
                \If {$tb>ta$}{
                    $G.nodes[u]['tend'] \gets 2.5$\;
            }

            }
                \KwRet{$G$}
                
		}
	}
    }
    }
	\caption{Diffuse the information in the network $G$ and return the updated graph with updated activated vertices in the graph $G$}
	\label{algo:diffu}
\end{algorithm}

\begin{algorithm}
	\DontPrintSemicolon
	\SetKwInOut{Input}{Input}\SetKwInOut{Output}{Output}
	\Input   {$\theta_1, \theta_2, G, S_A, \text{ and }S_B$.}
	\Output  {Return the set of  temporary activated vertices $T_A$.}
	\SetKwFunction{FMain}{\tempactive{}}
	\SetKwProg{Fn}{Function}{}
	\Fn{\FMain{$\theta_1, \theta_2, G, S_A, S_B$}} {
		\Begin{
        $T_A \gets \phi$\;
    \For{ $u \in V(G)$}{
        $t_a \gets 0$\;
        $t_b \gets 0$\;
        $t_c \gets 0$\;
        \If {$u$ is not active}{
        \For{ $v \in N_{in}[u]$}{
            \If {$G.nodes[v]['tend']=1.5$}{
                $t_a = t_a + W_{(v,u)}$\;
                }
            \ElseIf {$G.nodes[v]['tend']= 2.5$}{
                $t_b = t_b + W_{v,u}$\;
                }
            $t_c \gets t_a + t_b$\;
        }
         
        \If {$G.nodes[u]['tend']=2$}{
            \If {$t_b>= \theta_1$ or $ t_a >= \theta_1 + \theta_2$}{
                $T_A \gets T_A \cup \{u \}$\;
                }
            }
        \ElseIf{$G.nodes[u]['tend']=1$}{
            \If{$t_a \ge \theta_1 $ or $ t_b >= \theta_1 + \theta_2$}{
                $T_A \gets T_A \cup \{u \}$\;
                }
                }
        \ElseIf{$t_c >= \theta_1$}{ 
            $T_A \gets T_A \cup \{u \}$\;
            }
        
   % \KwRet{$tempA$}
                
    }
    }
    \KwRet{$T_A$}
    }
    }
	\caption{Find the set of the vertices $T_A$ those are temporary active.}
	\label{algo:tempAct}
\end{algorithm}

\section{\CSE{}}
\label{sec:a3}
The \CSE{} for the UDCIM problem finds the competitor seed set $S_B$ based on derived formulae from the page-rank score and the belonging community size of the vertex.  The formula is $\sum\limits_{u \in c} pr(u)* |c|$ where $pr(u)$ is the PageRank centrality score of the vertex $u$. The inputs for the algorithm are $G,S_A,k,\theta_1$, and $\theta_2$ and returns the seed set $S_B$ of size $k$. As usual, the genetic algorithm passes through four major stages to find the seed set $S_B$.
\begin{itemize}
    \item Create a population of size $p$ from the given graph $G$. The population size $p$ has $p$ individuals. An individual is a random $ k$-size subset of $V(G)$, where $k$ is the size of the seed set $S_B$ and $S_A$ or input parameter.
    
    \item Select two individuals from the population by the tournament selection method.
    \item Crossover takes two individuals and generates two offspring.
    \item Do a probabilistic mutation of these two individuals and update the new individuals.
    \item The new individuals are added to the population with the fitness score, and if the new individuals' fitness scores are better than at least one individual in our population, remove the two individuals with the lowest fitness scores to preserve the size  of the population at $p$.
\end{itemize}

 \begin{enumerate}
     \item \textbf{Population Creation:} Each individual in the population of size $k$ for the seed set $S_B$ is generated randomly.
     \item \textbf{Fitness Function:} As given below, the fitness function $f$ of $individual$ in the population is the difference between the number of final influenced vertices by $S_B$ and the number of final influenced vertices by $S_A$. $$f(individual)= \sigma(individual)-\sigma(S_A)$$ 

     \item \textbf{Selection:} We use the tournament selection method to balance exploration and exploitation.
     
     \item \textbf{Crossover:}  A one-point crossover is used to generate two offspring.  
     
     \item \textbf{Mutation:} Let $C$ be an array of communities of the graph $G$. The community array $C$ has a community $c$ only if at least one vertex $u\in c$ belongs to $\sigma(S_A)$. The score of each community $c\in C$, $S_C(c)$ is defined as $\sum\limits_{u \in c} pr(u)* |c|$ where $pr(u)$ is the PageRank centrality score of the vertex $u$. Sort the communities in decreasing order by calculated score $S_C(c)$. Traverse the decreasing order of communities and add the vertices to the list $P$. Sort the list $P$ in decreasing order of out-degree. For each  $u\in P$, if the selection probability of the vertex $u$ is under $0.5$, then the particular vertex $u$ is selected to mutate a randomly selected vertex from an individual. The 5\% vertices of the individual are randomly selected for mutation. %The mutation does more exploration in search space locally.  
 \end{enumerate}
 
%The algorithm is run $10$ multiple times and each run hs $100$ iterations.

\section{Result and Discussion}
\label{sec:result}
We use real-world $11$ datasets collected from different sources ~\cite{2015MarkNewMan,reza2009social,snapnets,neworkRepository}. The properties of the nine datasets out of $11$ are tabulated in \tableautorefname~\ref{tab:grp}. The medium sized nine datasets take \textbf{$3.6$ hours} for all algorithms. The large-sized datasets used in our results are \textit{soc-Epinions1} and \textit{com-dblp} with size $|V|=75,879$ and $|V|=3,17,080$ respectively. We propose two algorithms \PBE{} (\PBH{}) and \CSE{} (\CSH{}). The algorithms \PBH{} and \CSH{} are compared with \UDH{}. All algorithms are run for almost \textbf{$7$ hours} on the Kaggle platform with $GPU\;P100$. We compare our algorithms' performance on time and quality-wise with existing results proposed by~\citeauthor{working!}~\cite{working!}.

The algorithms ' results on a large-sized graph \textit{Directed Epinions social network}~\cite{snapnets} is shown in \tableautorefname~\ref{tb:lresutl}. The authors' algorithms run for almost $4$ hours but did not give output, but our algorithms \PBH{} and \CSH{} give output in $1$ hour and $17$ minutes. The results are tabulated in \tableautorefname~\ref{tb:lresutl}.

The largest dataset in our experiments is \textit{com-dblp} with size $0.3$ million vertices and $1$ million edges. Existing algorithms take more than $5$ hours without stopping, and our algorithms (\PBH{} and \CSH{}) converge within $3$ hours and $17$ minutes.

\begin{table}[htb]
    \centering
     \caption{Network properties of the data sets~\cite{2015MarkNewMan,reza2009social,snapnets,neworkRepository} used in this paper.}
     \scriptsize
    \begin{tabular}{p{2.5cm} p{1.5cm} p{1.5cm} p{1.5cm} p{2cm} p{1.5cm} p{1cm}}
    \hline
        Network & Nodes  & Edges & Density  & Avg-Degree& Avg-triangles & Avg-CC \\
        \hline
        \hline
        power  &  4941  &  6594  &  0.0005    &  2.66 &  0.39  &  0.08 \\ \hline
        %BlogCatalog  &  10312  &  333983  &  0.006    &  64.77 &  1631.69 &  0.46 \\ \hline
        lastfm  &  7624  &  27806  &  0.001    &  7.29 &  15.91 &  0.21 \\ \hline
        tvshow &  3892  &  17262  &  0.002    &  8.87  &  0.37 &  67.13 \\ \hline
        web-polblogs  &  643  &  2280  &  0.01    &  7.09 &  14.01  &  0.23  \\ \hline
        %FC  &  4039  &  88234  &  0.0108  &  1197.3335  &  43.691  &  0.606 \\ \hline
        %CA-HepTh  &  9877  &  25998  &  0.0005    &  5.26 &  8.60  &  0.47 \\ \hline
        %polblogs  &  1224  &  16718  &  0.02    &  27.31 &  247.65  &  0.32 \\ \hline
        %Karate  &  34  &  78  &  0.13    &  4.58 &  3.97 &  0.57 \\ \hline
        hamster  &  1858  &  12534    &  27.04 &  0.002  &  13.49  &  0.14 \\ \hline
        %musae-crocodile  &  11631  &  170918  &  0.01    &  29.39  &  160.80 &  0.33 \\ \hline
        facebook  &  4039  &  88234  &  0.01    &  43.69 &  1197.33  &  0.60 \\ \hline
        CA-GrQc  &  5242  &  14496  &  0.01    &  5.53 &  27.61  &  0.53 \\ \hline
        %Reed98  &  962  &  18812  &  0.04    &  39.11 &  302.92 &  0.31 \\ \hline
        %musae-squirrel &  5201  &  198493  &  0.0031    &  76.32 &  5534.86 &  0.42 \\ \hline
        %government &  7057  &  89455  &  0.01    &  25.35 &  222.69  &  0.41 \\ \hline
        musae-chameleon &  2277  &  31421  &  0.01   &  27.59  &  451.99 &  0.48 \\ \hline
        politician &  5908  &  41729  &  0.001    &  14.12 &  88.67  &  0.38 \\ \hline
        %CA-HepPh  &  12008  &  118521  &  0.01   &  19.74 &  839.06   &  0.61 \\ \hline
        %jazz  &  198  &  2742  &  0.14    &  27.69 &  271.19 &  0.61 \\ \hline
        %HU\_edges  &  47538  &  222887  &  0.0002    &  9.37 &  6.19 &  0.11 \\ \hline
        %mahandas  &  1258  &  7619  &  0.009    &  12.11 &  8.14  &  0.06 \\ \hline
        
    \end{tabular}
    \label{tab:grp}
\end{table}

%['soc-Epinions1', 23573, 22, 46283, 201.0, 166.5, 45452.5, 687.5, 36888, 14, 46286, 306.9, 72.0, 45977.5, 639.1]
%[['soc-Epinions1', 23573, 22, 46283, 201.0, 166.5, 45452.5, 687.5, 36888, 14(A), 46286(by B), 306.9(T), 72.0, 45977.5, 639.1]]

 \begin{table}
     \centering
     \begin{tabular}{ccccccc|}
     \hline
          \multirow{2}{*}{Dataset} & \multirow{2}{*}{Algo-name}  &\multicolumn{2}{c}{$|S_A|=30$}  & \multicolumn{2}{c}{$|S_A|=40$}\\
          \cline{3-6}
          & &$\sigma(S_A)$  & $\sigma(S_B)$ & $\sigma(S_A)$ & $\sigma(S_B)$ \\
          \hline
          \multirow{2}{*}{com-dblp}& \PBH{} & 1702  & 174543 & 819 & 175806 \\
          &\CSH{} &  94044.5 & 18391.4 & 3763.2 & 124512.0\\
          \hline
          \multirow{2}{*}{soc-Epinions1}& \PBH{} &22  & 46283 & 14 & 46286 \\
          &\CSH{}&  166.5 & 45452.3 & 72.2 & 45977.4\\
          
          \hline
     \end{tabular}
     \caption{The above algorithms are run on dataset \textit{soc-Eipinions1}~\cite{neworkRepository} of size $75,879$ vertices and $508837$ edges and \textit{com-dbpl}~\cite{snapnets} with size $3,17,080$ and edges around $1$ millions.}
     \label{tb:lresutl}
 \end{table}

%[['soc-Epinions1', 5525, 14, 45876, 82.9, 35.0, 37310.0, 889.8], ['com-dblp', 10, 2, 16458, 66.9, 10.0, 62208.0, 3478.7]]

\subsection{Quality-wise performance}
In \figureautorefname~\ref{fig:inf(B5)}, $|S_A|=30$, the number of influenced nodes by seed set $S_B$ is represented on $Y-axis$ and dataset names are shown on $X-axis$.

The objective of the \UDH{} problem is to maximize the number of influenced nodes by the seed set $S_B$ of size $30$. We show the  number $\sigma(S_B)-|S_B|$ in the plotted \figureautorefname~\ref{fig:inf(B5)}. For the same reason, the number of influenced nodes by $S_B$ is close to zero for the graph dataset \textit{musae-chameleon} by algorithm \UDH{}. % The Algorithm \PBH{} has a lower number for only one dataset \textit{lastfm}, but \CSH{} gives a better number for the same dataset.
The numbers of the algorithm \PBH{} are much better than the algorithm \UDH{}. For the \textit{facebook }, \textit{musae-chameleon }, and \textit{lastfm} datasets, the difference is quite large between \UDH{} and either of the \PBH{} and \CSH{} algorithms. 

When the seed set sizes $S_A$ are $40$ and  $50$  in \figureautorefname~\ref{fig:inf(B10)} and ~\ref{fig:inf(B20)} respectively, the gap between \UDH{} and our algorithms become large. For the dataset $lastfm$, when $|S_A|=40$,  \UDH{}  post comparative very few number of influenced nodes by $S_B$ while our algorithms give more than $1000$ number. The \CSE{} (\CSH{}) is somehow in between algorithms \UDH{} and \PBH{}.  Similarly, the trend for seed set size $50$ is preserved in \figureautorefname~\ref{fig:inf(B20)}.

The objective of the model is to maximize the influence of $S_B$ over $S_A$. The minimizing influence of $S_A$  and maximizing the influence of $S_B$ is also plotted in \figureautorefname~\ref{fig:inf(A5)}~\ref{fig:inf(A10)}~\ref{fig:inf(A20)}. In \figureautorefname~\ref{fig:inf(A5)}, $\sigma(S_A)$ is the number of influenced nodes by $S_A$. The blue bars represents the case for $\sigma(S_A)$ if $S_B= \phi$ and other bars represent the case $\sigma(S_A)$ if $S_B \ne \phi$. It means \UDH{}, \PBH{}, and \CSH{} select the seed set $S_B$ in such a way that it leads to minimizing the influenced nodes by $S_A$ and maximizing the influenced nodes by $S_B$. We measure how much $S_B$ can suppress the influence of $S_A$. The bars representing \PBH{} and \CSH{} algorithms are lower height than the blue bar. The gap is big between our algorithms and \UDH{} for the datasets \textit{musae-chameleon, web-polblogs, facebook}, and \textit{hamster}. In other \figureautorefname~\ref{fig:inf(A10)} and ~\ref{fig:inf(A20)}, we see the same trends.

\subsubsection{Reason behind performance over existing result}
\label{sec:result_qualtiy}
As the authors use the eigenvector centrality to find the centrality of the node. But we use the PageRank algorithm to measure the centrality of the nodes. The eigen vector centrality is less effective than the PageRank algorithm for directed graphs. Another thing, authors are also finding seed nodes of the set $S_B$ in an identical community where $S_A$ influences the nodes, and then based on the centrality score, the vertex is added as a seed node in the set $S_B$. Whereas, we only find the vertices $u$ such that there is a path from $u$ to $v \in \sigma(S_A)$, and a vertex with a high out-degree is added as a seed vertex in the set $S_B$. Because, inside a community, a global centrality score of the whole graph is less effective as compared to out-degree of the vertex. The out-degree captures the local structure of the graph in a community.  Similarly, in the \CSE{}, the rank of vertices is decided by multiplying the PageRank of the vertex by the size of the community the vertex belongs to. As we see in the \figureautorefname~\ref{fig:inf(B5)},~\ref{fig:inf(B10)}, and ~\ref{fig:inf(B20)} our algorithms give better results.   

\section{Runtime comparison}
\label{sec:result_runtime}
%The time complexity of recently proposed algorithm A1~\cite{working!} is $O(V'(V+E))$ where $V'$ is top $30\%$ vertices of $V$. But when the size of graph increased  
%In figures \figureautorefname~\ref{fig:time(5)} ,~\ref{fig:time(10)} and~\ref{fig:time(20)}, our algorithm is performing
The \UDH{} algorithm first calculates the topological interest(TI). To find TI for all vertices takes time $O(|V|)$. As the rank of the vertices is pre-computed. The major computation time is taken by \UDH{} to find the seed vertex out of $30\%$ of the vertices of the graph $G$. For each $u\in V$, the diffusion algorithm is run, and then the top $30\%$ vertices are selected.  Hence, the time complexity of the \UDH{} is $O(|V|(|V|+|E|))$ where $V$ and $E$ are the set of vertices and edges. On the other hand, our algorithm \PBH{} only finds all the vertices $v \in V$ such that there exists a directed path to any $u\in \sigma(S_A)$ from $v$, $u$ and $v$ belong to the same community.

As we explained in section ~\ref{sec:a2}, the time complexity of \PBH{} is $O(|C| (|V| + |E|))$, where $C$ is set of communities in the graph $G$. The effect is clearly visible on the computation time of the \CSH{} and \PBH{} algorithms on the \figureautorefname~\ref{fig:time(10)}~\ref{fig:time(20)}~\ref{fig:time(5)}.

\begin{comment}

\end{comment}

\pgfplotstableread[col sep=comma,]{data.csv}\datatable
\begin{figure}
    \centering
\begin{tikzpicture}
\begin{axis} [
    height=7cm,
    x post scale=1.5,
    ybar,
    bar width=6pt,
    ylabel={$\sigma(S_B)$},
    xtick=data,
    xticklabels from table={\datatable}{name},
    ymajorgrids,
    xmajorgrids,
    x tick label style={rotate=90,/pgf/number format/1000 sep=3pt},
    legend pos=north west]

%\addplot table [x=name, y=inf(A'15),]{\datatable};
%\addplot table [x=name, y=inf(A15),]{\datatable};
%\addplot table [x=name, y=inf(B15),]{\datatable};
\addplot table [x expr=\coordindex, y=inf(B15)]{\datatable};

%\addplot table [x=name, y=inf(A25),]{\datatable};
\addplot table [x expr=\coordindex, y=inf(B25)] {\datatable};

%\addplot table [x=name, y=inf(A35),]{\datatable};
\addplot table [x expr=\coordindex, y=inf(B35)] {\datatable};

\legend{\UDH{}, \PBH{},\CSH{}}
\end{axis}

\end{tikzpicture}

\caption{The seed set $|S_A|=30$, $Y$-axis shows the number of influenced nodes by $S_B$ ,i.e $\sigma(S_B)$.}
    \label{fig:inf(B5)}
\end{figure}
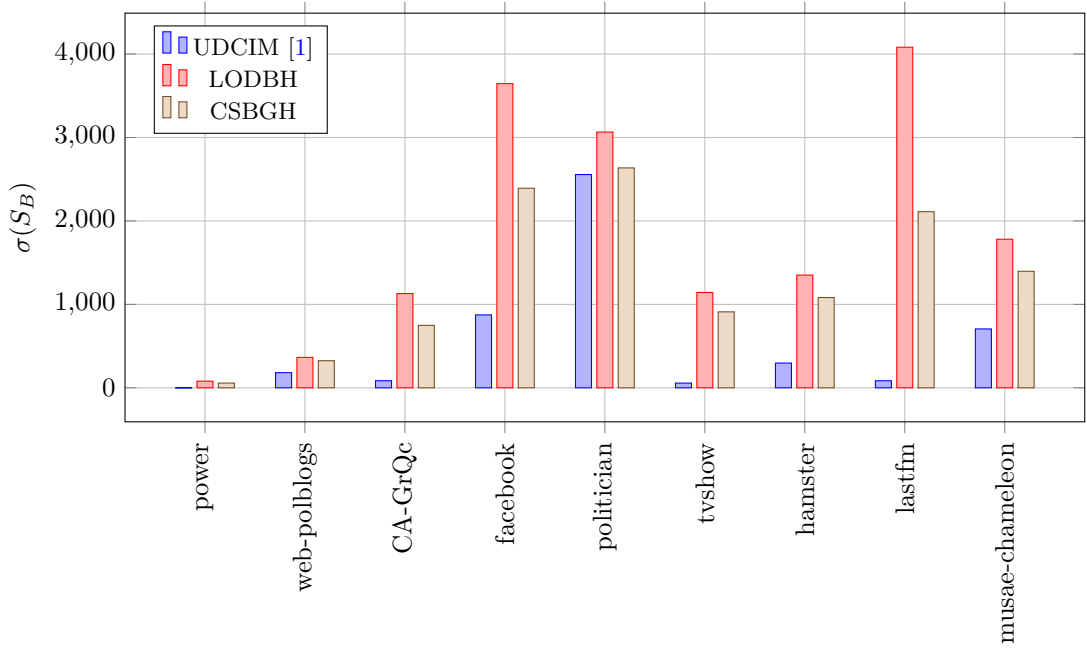

\begin{figure}
\centering
\begin{tikzpicture}
\begin{axis} [
    height=7cm,
    x post scale=1.5,
    ybar,
    bar width=6pt,
    ylabel={$\sigma(S_A)$},
    xtick=data,
    xticklabels from table={\datatable}{name},
    ymajorgrids,
    xmajorgrids,
    x tick label style={rotate=90,/pgf/number format/1000 sep=3pt},
    legend pos=north west]

\addplot table [x expr=\coordindex, y=inf(A'15)]{\datatable};
\addplot table [x expr=\coordindex, y=inf(A15)]{\datatable};
%\addplot table [x=name, y=inf(B15)]{\datatable};

\addplot table [x expr=\coordindex, y=inf(A25)]{\datatable};
%\addplot table [x=name, y=inf(B25)]{\datatable};

\addplot table [x expr=\coordindex, y=inf(A35)]{\datatable};
%\addplot table [x=name, y=inf(B35)]{\datatable};

\legend{$\sigma(S_A)$ \& $S_B=\phi$,\UDH{} \& $S_B\ne\phi$, \PBH{} \& $S_B\ne\phi$,\CSH{} \& $S_B\ne\phi$}
\end{axis}
\end{tikzpicture}

\caption{The seed set $|S_A|=30$, the blue bars indicate the diffusion of $S_A$ when $S_B=\phi$. The rest of the other color bars are the diffusion of $S_A$ after finding the seed set $S_B$ by running the corresponding algorithms (\UDH{}, \PBH{}, and \CSH{}).}
    \label{fig:inf(A5)}
\end{figure}
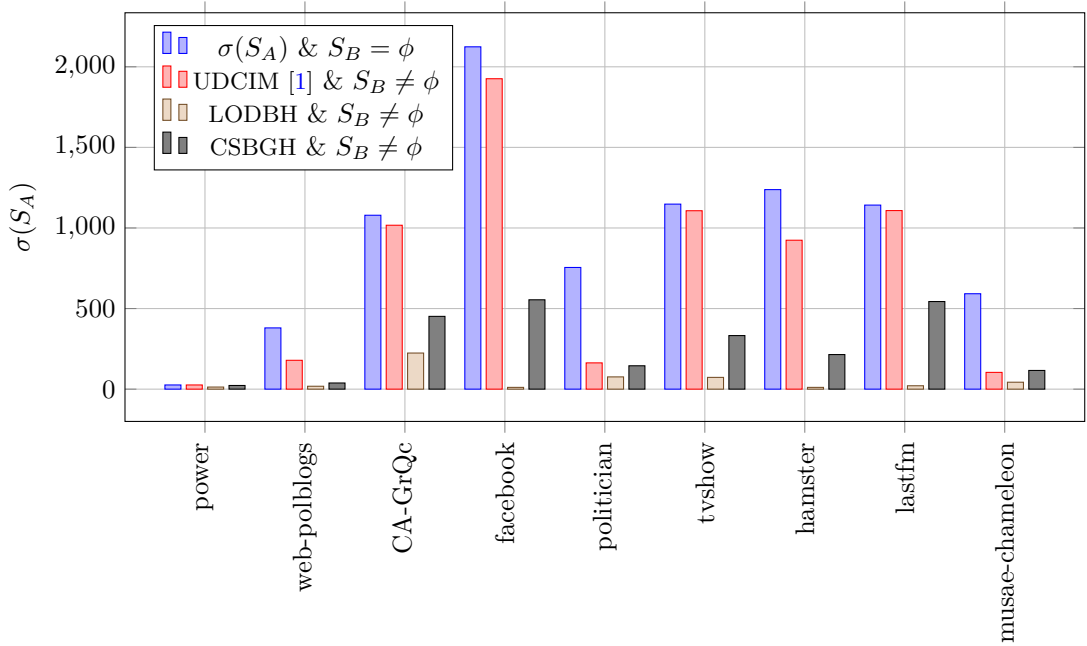

\begin{figure}
\centering
\begin{tikzpicture}

\begin{axis} [
    height=7cm,
    x post scale=1.5,
    ybar,
    bar width=6pt,
    ylabel={$\sigma(S_B)$},
    xtick=data,
    xticklabels from table={\datatable}{name},
    ymajorgrids,
    xmajorgrids,
    x tick label style={rotate=90,/pgf/number format/1000 sep=3pt},
    legend pos=north west]

%\addplot table [x=name, y=inf(A'110)]{\datatable};
%\addplot table [x=name, y=inf(A15)]{\datatable};
\addplot table [x expr=\coordindex, y=inf(B110)]{\datatable};

%\addplot table [x expr=\coordindex,, y=inf(A25)]{\datatable};
\addplot table [x expr=\coordindex, y=inf(B210)]{\datatable};

%\addplot table [x expr=\coordindex, y=inf(A35)]{\datatable};
\addplot table [x expr=\coordindex, y=inf(B310)]{\datatable};

\legend{\UDH{}, \PBH{},\CSH{}}
\end{axis}
\end{tikzpicture}

\caption{The seed set $|S_A|=40$, $Y$-axis shows the number of influenced nodes by $S_B$.}
    \label{fig:inf(B10)}
\end{figure}
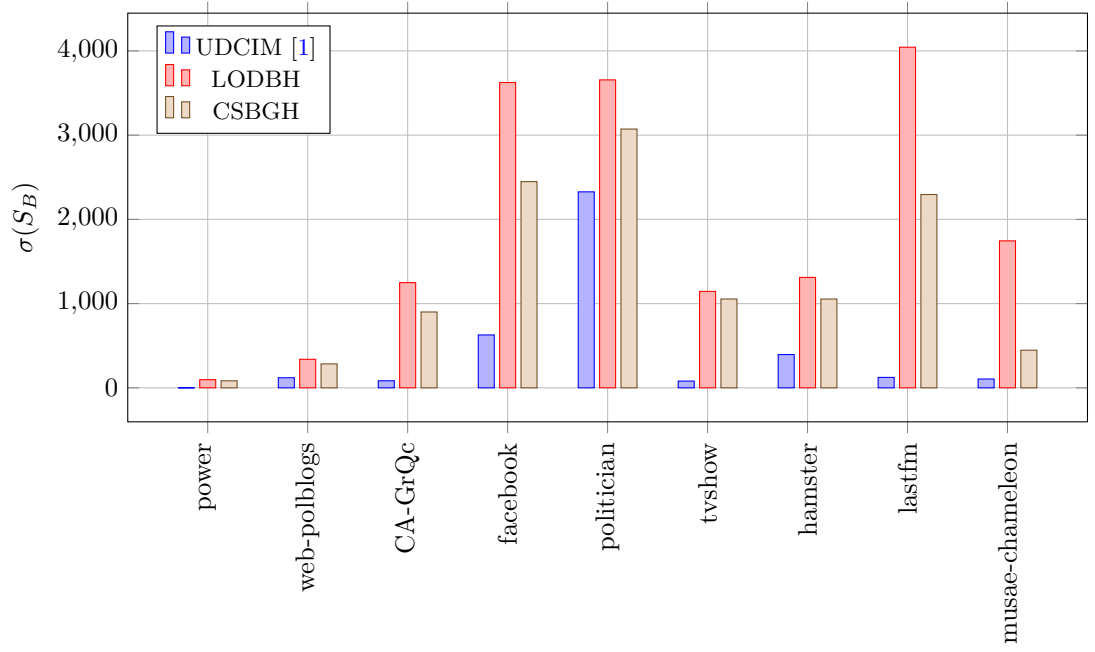

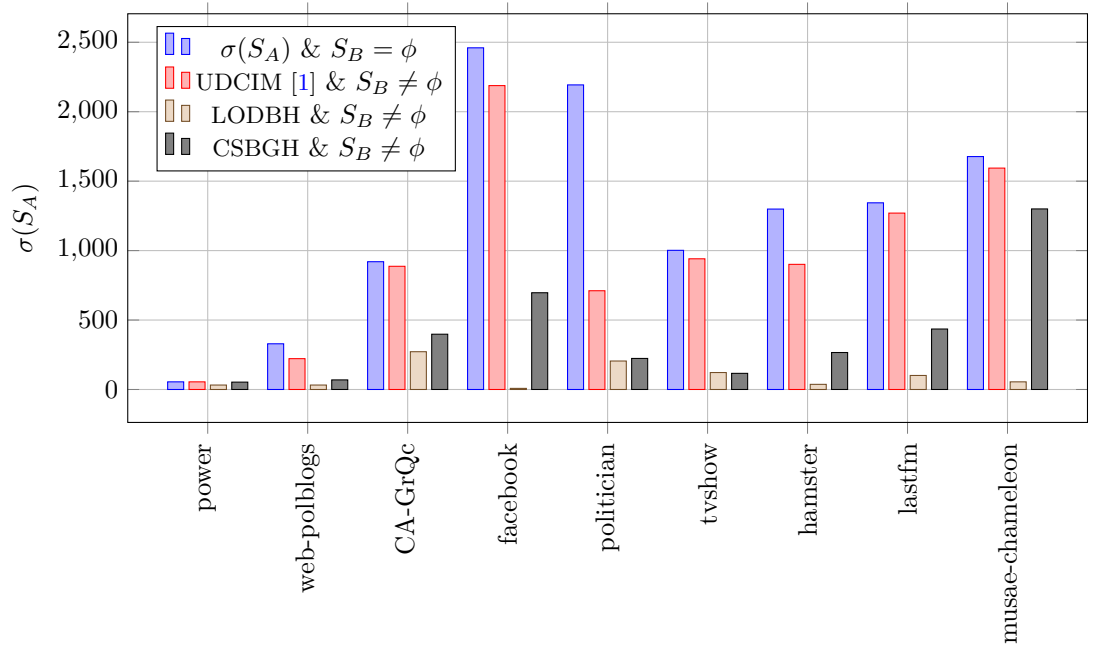
\begin{figure}
\centering
\begin{tikzpicture}

\begin{axis} [
    height=7cm,
    x post scale=1.5,
    ybar,
    bar width=6pt,
    ylabel={$\sigma(S_A)$},
    xtick=data,
    xticklabels from table={\datatable}{name},
    ymajorgrids,
    xmajorgrids,
    x tick label style={rotate=90,/pgf/number format/1000 sep=3pt},
    legend pos=north west]

\addplot table [x expr=\coordindex, y=inf(A'110)]{\datatable};
\addplot table [x expr=\coordindex, y=inf(A110)]{\datatable};
%\addplot table [x expr=\coordindex, y=inf(B15)]{\datatable};

\addplot table [x expr=\coordindex, y=inf(A210)]{\datatable};
%\addplot table [x expr=\coordindex, y=inf(B25)]{\datatable};

\addplot table [x expr=\coordindex, y=inf(A310)]{\datatable};
%\addplot table [x expr=\coordindex, y=inf(B35)]{\datatable};

\legend{$\sigma(S_A)$ \& $S_B=\phi$,\UDH{} \& $S_B\ne\phi$, \PBH{} \& $S_B\ne\phi$,\CSH{} \& $S_B\ne\phi$}
\end{axis}
\end{tikzpicture}

\caption{The seed set $|S_A|=40$, the blue bars indicate the diffusion of $S_A$ when $S_B=\phi$. The rest of the other color bars are the diffusion of $S_A$ after finding the seed set $S_B$ by running the corresponding algorithms (\UDH{}, \PBH{}, and \CSH{}).}
    \label{fig:inf(A10)}
\end{figure}

\begin{figure}
\centering
\begin{tikzpicture}

\begin{axis} [
    height=7cm,
    x post scale=1.5,
    ybar,
    bar width=6pt,
    ylabel={$\sigma(S_B)$},
    xtick=data,
    xticklabels from table={\datatable}{name},
    ymajorgrids,
    xmajorgrids,
    x tick label style={rotate=90,/pgf/number format/1000 sep=3pt},
    legend pos=north west]

%\addplot table [x expr=\coordindex, y=inf(A'120)]{\datatable};
%\addplot table [x expr=\coordindex, y=inf(A15)]{\datatable};
\addplot table [x expr=\coordindex, y=inf(B120)]{\datatable};

%\addplot table [x expr=\coordindex, y=inf(A25)]{\datatable};
\addplot table [x expr=\coordindex, y=inf(B220)]{\datatable};

%\addplot table [x expr=\coordindex, y=inf(A35)]{\datatable};
\addplot table [x expr=\coordindex, y=inf(B320)]{\datatable};

\legend{\UDH{}, \PBH{},\CSH{}}
\end{axis}
\end{tikzpicture}

\caption{The seed set $|S_A|=50$, $Y$-axis shows the number of influenced nodes by $S_B$.}
    \label{fig:inf(B20)}
\end{figure}
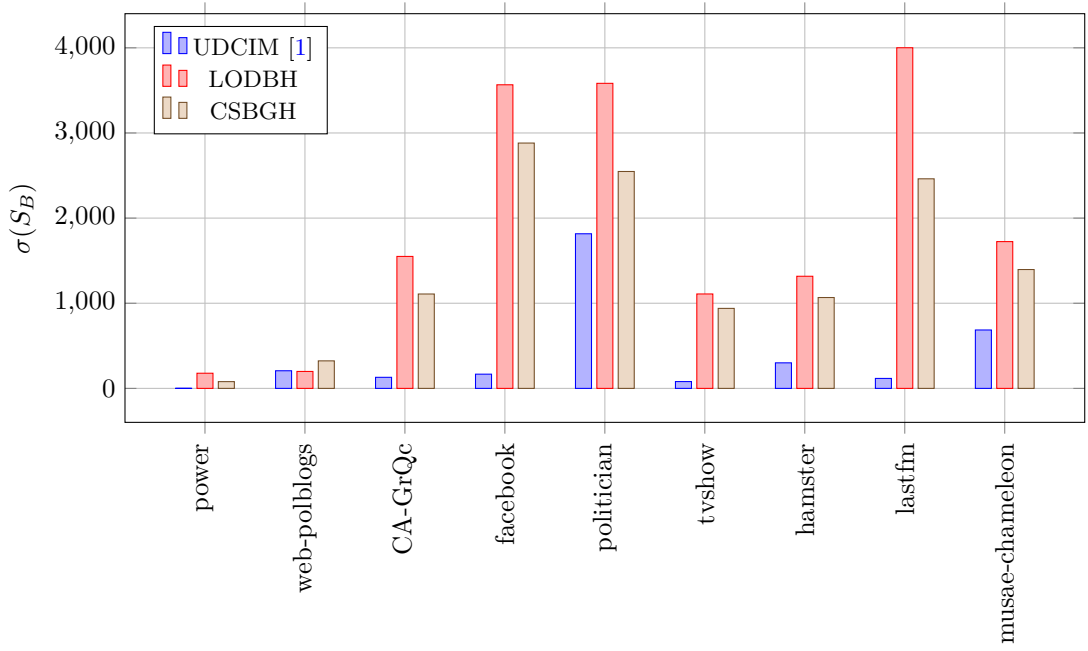

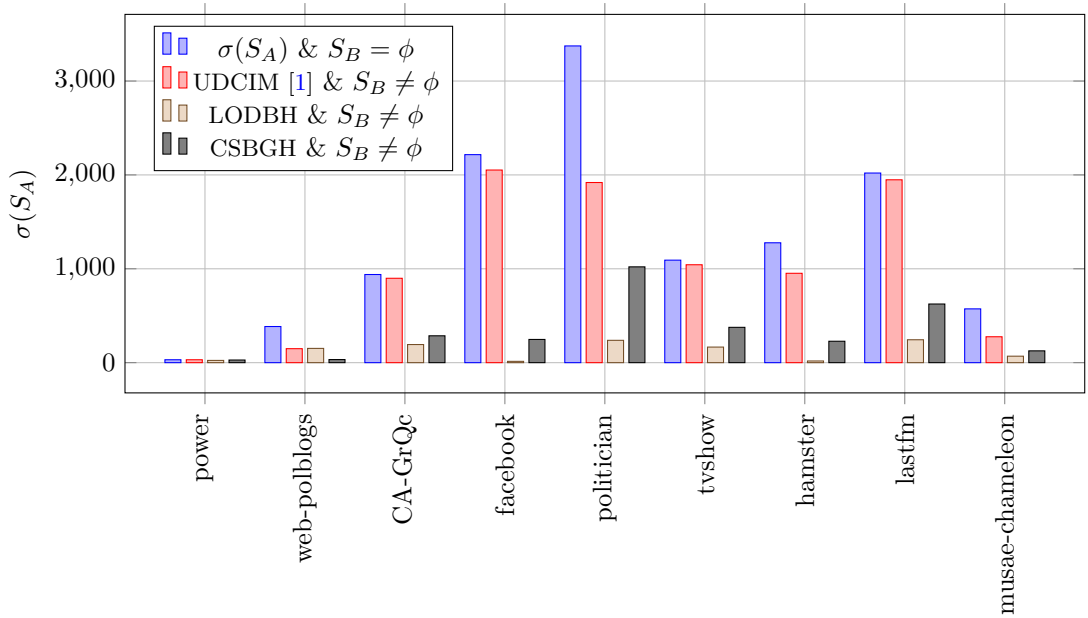
\begin{figure}
    \centering
\begin{tikzpicture}

\begin{axis} [
    height=6.6cm,
    x post scale=1.5,
    ybar,
    bar width=6pt,
    ylabel={$\sigma(S_A)$},
    xtick=data,
    xticklabels from table={\datatable}{name},
    ymajorgrids,
    xmajorgrids,
    x tick label style={rotate=90,/pgf/number format/1000 sep=3pt},
    legend pos=north west]

\addplot table [x expr=\coordindex, y=inf(A'120)]{\datatable};

\addplot table [x expr=\coordindex, y=inf(A120)]{\datatable};
%\addplot table [x expr=\coordindex, y=inf(B15)]{\datatable};

\addplot table [x expr=\coordindex, y=inf(A220)]{\datatable};
%\addplot table [x expr=\coordindex, y=inf(B25)]{\datatable};

\addplot table [x expr=\coordindex, y=inf(A320)]{\datatable};
%\addplot table [x expr=\coordindex, y=inf(B35)]{\datatable};

\legend{$\sigma(S_A)$ \& $S_B=\phi$,\UDH{} \& $S_B\ne\phi$, \PBH{} \& $S_B\ne\phi$,\CSH{} \& $S_B\ne\phi$}
\end{axis}

\end{tikzpicture}

\caption{The seed set $|S_A|=50$, the blue bars indicate the diffusion of $S_A$ when $S_B=\phi$. The rest of the other color bars are the diffusion of $S_A$ after finding the seed set $S_B$ by running the corresponding algorithms (\UDH{}, \PBH{}, and \CSH{}).  }
    \label{fig:inf(A20)}
\end{figure}

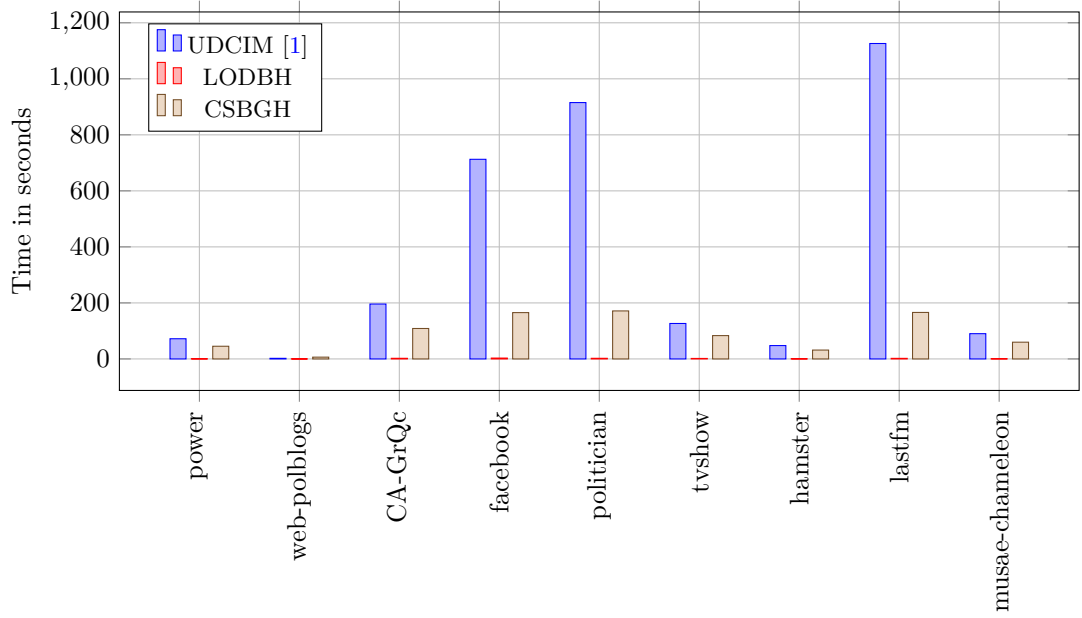
\begin{figure}
    \centering
\begin{tikzpicture}
\begin{axis} [
    height=6.6cm,
    x post scale=1.5,
    ybar,
    bar width=6pt,
    ylabel={Time in seconds},
    xtick=data,
    xticklabels from table={\datatable}{name},
    ymajorgrids,
    xmajorgrids,
    x tick label style={rotate=90,/pgf/number format/1000 sep=3pt},
    legend pos=north west]

%\addplot table [x expr=\coordindex, y=inf(A'120)]{\datatable};
\addplot table [x expr=\coordindex, y=time15]{\datatable};
%\addplot table [x expr=\coordindex, y=inf(B15)]{\datatable};

\addplot table [x expr=\coordindex, y=time25]{\datatable};
%\addplot table [x expr=\coordindex, y=inf(B25)]{\datatable};

\addplot table [x expr=\coordindex, y=time35]{\datatable};
%\addplot table [x expr=\coordindex, y=inf(B35)]{\datatable};

\legend{\UDH{}, \PBH{},\CSH{}}
\end{axis}

\end{tikzpicture}

\caption{When the seed size $|S_A|= 30$, \PBH{} and \CSH{} reduce the time compared to \UDH{}}
\label{fig:time(5)}

\end{figure}

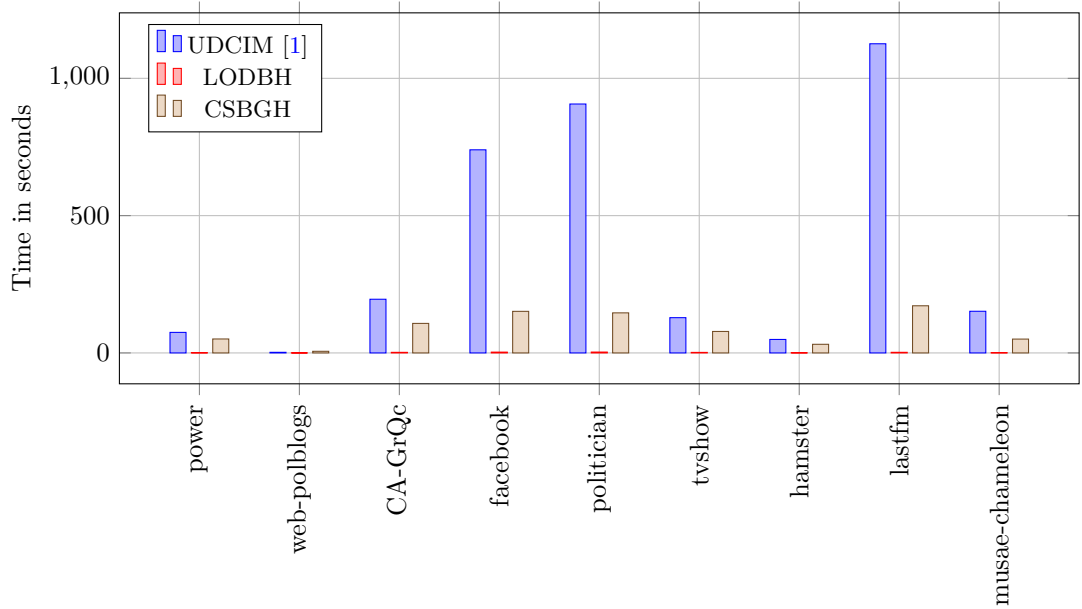
\begin{figure}
\centering
\begin{tikzpicture}
\begin{axis} [
    height=6.5cm,
    x post scale=1.5,
    ybar,
    bar width=6pt,
    ylabel={Time in seconds},
    xtick=data,
    xticklabels from table={\datatable}{name},
    ymajorgrids,
    xmajorgrids,
    x tick label style={rotate=90,/pgf/number format/1000 sep=1pt},
    legend pos=north west]

%\addplot table [x expr=\coordindex, y=inf(A'120)]{\datatable};
\addplot table [x expr=\coordindex, y=time110]{\datatable};
%\addplot table [x expr=\coordindex, y=inf(B15)]{\datatable};

\addplot table [x expr=\coordindex, y=time210]{\datatable};
%\addplot table [x expr=\coordindex, y=inf(B25)]{\datatable};

\addplot table [x expr=\coordindex, y=time310]{\datatable};
%\addplot table [x expr=\coordindex, y=inf(B35)]{\datatable};

\legend{\UDH{}, \PBH{},\CSH{}}
\end{axis}

\end{tikzpicture}

\caption{When the seed size $|S_A|= 40$, \PBH{} and \CSH{} reduce the time compared to \UDH{}}
\label{fig:time(10)}

\end{figure}
%\end{figure}

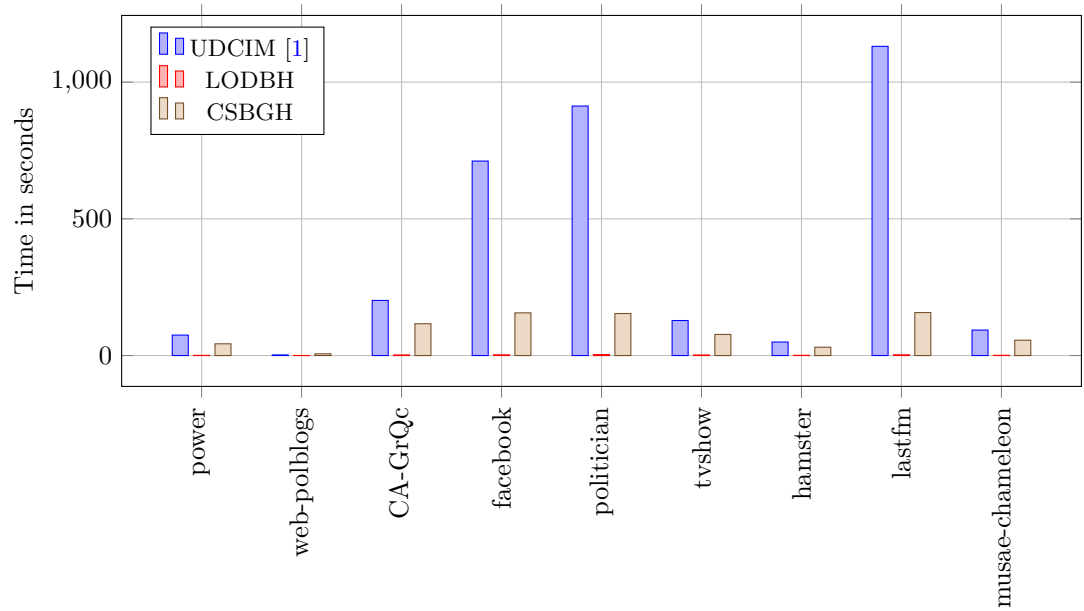
\begin{figure}
\centering
\begin{tikzpicture}
\begin{axis} [
    height=6.5cm,
    x post scale=1.5,
    ybar,
    bar width=6pt,
    ylabel={Time in seconds},
    xtick=data,
    xticklabels from table={\datatable}{name},
    ymajorgrids,
    xmajorgrids,
    x tick label style={rotate=90,/pgf/number format/1000 sep=3pt},
    legend pos=north west]

%\addplot table [x expr=\coordindex, y=inf(A'120)]{\datatable};
\addplot table [x expr=\coordindex, y=time120]{\datatable};
%\addplot table [x expr=\coordindex, y=inf(B15)]{\datatable};

\addplot table [x expr=\coordindex, y=time220]{\datatable};
%\addplot table [x expr=\coordindex, y=inf(B25)]{\datatable};

\addplot table [x expr=\coordindex, y=time320]{\datatable};
%\addplot table [x expr=\coordindex, y=inf(B35)]{\datatable};

\legend{\UDH{}, \PBH{},\CSH{}}
\end{axis}

\end{tikzpicture}
\caption{When the seed size $|S_A|= 50$, \PBH{} and \CSH{} reduce the time compared to \UDH{}}
\label{fig:time(20)}
\end{figure}

\clearpage

\section{Conclusion}
\label{sec:cnc}
The influence maximization in social networks is the center of attraction for researchers. In a competitive world, lots of information of the same type propagates in the social networks simultaneously. The influence of the intended information is maximized in the networks by developing a biased algorithm. We discuss the user-driven competitive influence maximization (\UDH{}) problem in social networks.

In this article, the LP-formulation of \UDH{} and the implementation of the LP-formulated problem of \UDH{} by using Gurobi Solver~\cite{gurobi} are discussed. For small datasets, the results between Gurobi~\cite{gurobi} and others are compared, where Gurobi~\cite{gurobi} outperforms the others in maximizing the influence for the seed set $S_B$. We also present one heuristic and one genetic algorithm. The heuristic algorithms (\PBH{} and \CSH{}) provides a better result as compared with the existing result, in terms of time and quality, as shown in \figureautorefname~\ref{fig:time(10)} and ~\ref{fig:inf(A10)}, respectively. To check the scalability of the algorithms, the experiments on large datasets \textit{soc-Epillion1} and \textit{com-dblp} are also carried out. The algorithm \UDH{} does not stop running till $4$ hours, whereas our algorithms return the results within $1$ hours and $17$ minutes for the dataset \textit{soc-Epinion1}. The largest dataset in our experiments is \textit{com-dblp} with size $0.3$ million vertices and $1$ million edges. Existing algorithms take more than $5$ hours without stopping, and both our algorithms (\PBH{} and \CSH{}) converge within $3$ hours and $17$ minutes. We discuss the reason for the faster computation of our algorithms and the existing ones. Similarly, we also discuss the strong arguments on quality-wise performance of our algorithms. The \UDH{} problem is still an open problem for an undirected graph and for trees to check whether the problem is solvable in polynomial time or not.

\bibliography{sn-bibliography}% common bib file
%% if required, the content of .bbl file can be included here once bbl is generated
%%\input sn-article.bbl

\end{document}